\documentclass[nonacm=true,sigconf]{acmart}
\AtBeginDocument{%
  \providecommand\BibTeX{{%
    \normalfont B\kern-0.5em{\scshape i\kern-0.25em b}\kern-0.8em\TeX}}}

\usepackage{booktabs,xspace,csquotes,tabularx,multirow,balance}

\newcommand{\etal}{et~al.\xspace}
\newcommand{\ie}{i.e.,\ }

\newcommand{\icwsm}{0}   
\newcommand{\para}[1]{{\vspace{7pt} \bf \noindent #1 \hspace{7pt}}}

\begin{document}

\title{Algorithms that ``Don't See Color'': Measuring Biases in Lookalike and Special Ad Audiences}

\author{Piotr Sapiezynski}
\affiliation{
    \institution{Northeastern University}
    \city{Boston}
    \state{MA}
    \country{USA}
}
\email{p.sapiezynski@northeastern.edu}

\author{Avijit Ghosh}
\affiliation{
    \institution{Northeastern University}
    \city{Boston}
    \state{MA}
    \country{USA}
}
\email{ghosh.a@northeastern.edu}

\author{Levi Kaplan}
\affiliation{
    \institution{Northeastern University}
    \city{Boston}
    \state{MA}
    \country{USA}
}
\email{kaplan.l@northeastern.edu}

\author{Aaron Rieke}
\affiliation{
    \institution{Upturn}
    \city{Washington}
    \state{DC}
    \country{USA}
}
\email{aaron@upturn.org}

\author{Alan Mislove}
\affiliation{
    \institution{Northeastern University}
    \city{Boston}
    \state{MA}
    \country{USA}
}
\email{amislove@ccs.neu.edu}


%!TEX root = main_icwsm.tex
\begin{abstract}

Researchers and journalists have repeatedly shown that algorithms commonly used in domains such as credit, employment, healthcare, or criminal justice can have discriminatory effects.
Some organizations have tried to mitigate these effects by simply removing sensitive features from an algorithm's inputs. 
In this paper, we explore the limits of this approach using a unique opportunity. 
In 2019, Facebook agreed to settle a lawsuit by removing certain sensitive features from inputs of an algorithm that identifies users similar to those provided by an advertiser for ad targeting, making both the modified and unmodified versions of the algorithm available to advertisers.
We develop methodologies to measure biases along the lines of gender, age, and race in the audiences created by this modified algorithm, relative to the unmodified one.
Our results provide experimental proof that merely removing demographic features from a real-world algorithmic system's inputs can fail to prevent biased outputs.
As a result, organizations using algorithms to help mediate access to important life opportunities should consider other approaches to mitigating discriminatory effects.
\end{abstract}

% %%
% %% The code below is generated by the tool at http://dl.acm.org/ccs.cfm.
% %% Please copy and paste the code instead of the example below.
% <concept_id>10003456.10003462</concept_id>
% <concept_desc>Social and professional topics~Computing / technology policy</concept_desc>
% <concept_id>10010405.10010481.10010488</concept_id>
% <concept_desc>Applied computing~Marketing</concept_desc>
% <concept_significance>300</concept_significance>

% \ccsdesc[500]{Social and professional topics~Computing / technology policy}
% \ccsdesc[300]{Applied computing~Marketing}

% %%
% %% Keywords. The author(s) should pick words that accurately describe
% %% the work being presented. Separate the keywords with commas.


\maketitle

%!TEX root = main_icwsm.tex
\section{Introduction}
Organizations use algorithmic models\footnote{Throughout this paper, we refer to a large class of algorithmic models using the now-common term ``algorithms'', especially those created through statistical modeling and machine learning.} (``algorithms'') in a variety of important domains, including healthcare~\cite{obermeyer-2019-dissecting}, credit~\cite{fourcade-2013-classification}, employment~\cite{hannak-2017-bias,chen-2018-chi}, and content distribution~\cite{ali-2019-discrimination}.
Unfortunately, these algorithms have been shown to sometimes have discriminatory effects that can often be challenging to detect, measure, and articulate.
%
% For example, in the context of criminal justice, ProPublica showed that the COMPAS risk-assessment tool used by judges to help make bail decisions was more likely to mislabel Black defendants as future criminals~\cite{courtland-2018-bias}, when compared to white defendants.
% %
% Similarly, facial recognition algorithms have been shown to perform significantly worse for Black women~\cite{buolamwini-2018-gender}.
% %
% Facebook, in its quest to show relevant ads to users, diverts the delivery of employment and housing ads away from some demographic groups, even when an advertiser is trying to reach a broad, diverse audience~\cite{ali-2019-discrimination, imana-2021-www}. 
Some have proposed mitigating discriminatory effects by removing demographic features from an algorithm's inputs.
%
%For an algorithm that screens loan applications, one could ensure that the algorithm is not provided with the sex, race, age, or any other protected characteristic of the input.
%
For example, in 2019 the U.S. Department of Housing and Urban Development (HUD) proposed a rule that considered applying this approach to housing discrimination~\cite{hud-2019-implementation}.
Because algorithms can effectively use omitted demographic features by combining other inputs that are each {\em correlated} with those features~\cite{barocas-2019-fairmlbook}, such a rule could nullify any protection from discriminatory effects. 
This is particularly true in large-scale machine learning (ML) systems, which can take as input thousands or even millions of features~\cite{blass-2019-algorithmic}.

In this paper, we leverage a unique opportunity created by a recent lawsuit settlement involving Facebook's advertising platform to explore the limits of this approach.
Specifically, we examine Facebook's {\em Lookalike Audiences}
\if \icwsm
~\cite{FacebookLookalikeAudience}
\fi
targeting tool, which takes a list of Facebook users provided by an advertiser (called the {\em source audience}) and creates a new audience of users who share ``common qualities'' with those in the source audience.
In March 2018, the National Fair Housing Alliance (NFHA) and others sued~\cite{FacebookCivilLawsuit} Facebook over violations of the Fair Housing Act (FHA).
%, specifically Title 24 CFR \S\ 100.75 {\em Discriminatory advertisements, statements and notices}. 
%
When the case was settled in March 2019, Facebook agreed to modify the functionality of Lookalike Audiences when used to target housing, credit, and employment ads.
In brief, Facebook created the {\em Special Ad Audiences} tool,
\if \icwsm
~\cite{FacebookSpecialAdAudience}
\fi which works like Lookalike Audiences, except its algorithm does {\em not} consider users' age, gender, relationship status, religious views, school, political views, interests, or zip code when detecting common qualities.
\if \icwsm
~\cite{FacebookSpecialAdCategory}
\fi

%The co-existence of the Lookalike Audiences and Special Ad Audiences tools offers a chance to compare how complex ML systems perform on large-scale, real-world data both when provided with protected attributes and when not.
%
We seek to learn whether the Special Ad Audience algorithm (which is not provided with certain demographic features) actually produces significantly less {\em skewed} audiences than the Lookalike Audience algorithm (which is). 
In other words, when provided with a source audience that skews heavily toward one demographic group over another, to what extent do each of these tools reproduce that skew?
We focus on skews along demographic features named in the settlement, enabling us to examine whether simply removing the protected features as input to an algorithm is sufficient eliminate skew along those features.
%Put simply, {\em do Special Ad Audiences actually produce less discriminatory outcomes than Lookalike Audiences?}
To do so, we develop a methodology to examine the delivery of the same ads when using the two types of audiences, measuring the skew along the lines of gender, age, and race.

We show that our Special Ad audiences\footnote{Throughout the paper, we use ``Lookalike Audience'' or ``Special Ad Audience'' to refer to the general tools provided by Facebook, and  ``Lookalike audience'' or ``Special Ad audience'' to refer to a particular audience.} are skewed to almost the same degree as Lookalike audiences, with many of the results being statistically indistinguishable.
For example, when using a source audience that is all women, our Lookalike audience-targeted ad delivered to 96.1\% women, while Special Ad audience-targeted ad delivered to 91.2\% women.
We also provide evidence indicating that both Lookalike and Special Ad audiences carry---to a certain extent---the biases of the source audience in terms of race and political affiliation.

To underscore the real-world impact of these results, we place ads as an employer who is seeking to find candidates ``similar to'' to their current workforce.
Using a source audience consisting of Facebook employees
%---created via randomly generated {\tt @fb.com} emails---
we find that the resulting Special Ad audience skews heavily towards 25--34-year-old men.
We also confirm that previous findings on how Facebook's delivery mechanisms can cause {\em further} skews in who is shown ads hold for Special Ad Audiences.
%
% Our ad for artificial intelligence jobs delivers mostly to young men, while our ad for supermarket jobs delivers mostly to middle-aged women, despite targeting the same gender- and age-balanced Special Ad audience.
% \end{packed_itemize}
%

Taken together, our results show that simply removing demographic features from the inputs of a large-scale, real-world algorithm will not always suffice to meaningfully change its outputs.
%
%The answer is negative: the bias that exists when providing the algorithm with the features is almost precisely the same as if those features are not provided, as the algorithm is provided with numerous other inputs that are each correlated in various ways with the protected features.
%
%To the extent that the protected features are predictive, the algorithm is able to derive and use it just as effectively.
%
At the same time, this work presents a methodology by which other algorithms could be studied.

To be clear, we are not claiming---and do not believe---that Facebook has {\em incorrectly} implemented Special Ad Audiences, or is in violation of its settlement agreement. 
Rather, the findings in this paper are a natural result of how complex algorithmic systems work in practice.

%The remainder of this paper is organized as follows:
%
%Section~\ref{sec:background} provides background on Facebook's ad targeting tools and related work.
%
%Section~\ref{sec:methodology} introduces our methodology and Section~\ref{sec:results} presents our results.
%
%Section~\ref{sec:discussion} provides a concluding discussion.
\para{Ethics}
%We took careful consideration of ethics when conducting the research in this paper.
%
The research has been reviewed by our Institutional Review Board and marked as exempt.
Further, we minimized harm to Facebook users by only running ``real'' ads, \ie if a user clicked on one of our ads, they were presented with a real-world site relevant to content the ad.
We did not have any direct interaction with the users who were shown our ad, and did not collect any of their information. 
Finally, we minimized harm to Facebook by running and paying for our ads just like any other advertiser, as well as flagging them as employment ads whenever applicable.  
%
% When we ran employment ads, we flagged them as such during the ad creation flow.

%!TEX root = main.tex
% \begin{figure*}
%      \vskip-0.15in
%         \includegraphics[width=0.485\textwidth]{figures/screenshot-lookalike}
%         \includegraphics[width=0.485\textwidth]{figures/screenshot-special}
%         \caption{Screenshots of creation process for both Lookalike Audiences (left) and Special Ad Audiences (right).  Both of them are the same from the advertiser's perspective:  The advertiser first selects a source Custom audience, then selects a target country, and finally selects the fraction of that country's users to include in the new audience.}\label{fig:creation}
% \end{figure*}

\section{Background}\label{sec:background}
In this section, we provide background on Facebook's ad targeting tools and overview related work.

\subsection{Facebook's ad targeting tools}
Facebook provides a range of {\em targeting} tools to help advertisers select an {\em audience} of users who will be eligible to see their ads.
For example, advertisers can select users through combinations of {\em targeting attributes},~\if \icwsm ~\cite{FacebookAdAttributes}\fi  including over 1,000 demographic, behavioral, and interest-based features.

More germane to this paper and its methods, Facebook also offers a number of other, more advanced targeting tools.
One such tool is {\em Custom Audiences},\if \icwsm~\cite{FacebookCustomAudienceDef}\fi~which allows advertisers indicate individual users that they wish to include in an audience.
To use Custom Audiences, an advertiser uploads a list of personally identifiable information (PII), potentially including names, email addresses, phone numbers, dates of birth, and mobile identifiers. \if \icwsm~\cite{FacebookCustomAudiencePII}\fi 
Facebook then compares those identifiers against its database of active users, and lets the advertiser include matched users in their target audience.

Another tool is {\em Lookalike Audiences}, \if \icwsm~\cite{FacebookLookalikeAudience}\fi  which creates an audience of users who share ``common qualities'' with users in a Custom audience provided by the advertiser (called the {\em source audience}).
The exact input qualities used by the algorithm in creating these audiences are not known and the documentation lists only two examples: demographic information and interests.
%
%
%The motivation behind Lookalike Audiences is to allow advertisers to use Facebook's platform to find potential future customers (i.e., users who are similar to their current customers).
% %
% A screenshot of Facebook's advertiser interface is shown in Figure~\ref{fig:creation} (left); the advertiser must select the country where they wish Facebook to select users from (``Audience Location'') and then must select the fraction of that country's population to include in the new Lookalike Audience (ranging from 1\% to 10\%). 
%
Prior work has demonstrated that Lookalike Audiences can reproduce demographic skews present in source audiences~\cite{speicher-2018-targeted}.

\subsection{Special Ad Audiences}
%Facebook has recently been criticized~\cite{FacebookHousingDiscrimination} over their targeting mechanisms, as some of these mechanisms enable malicious advertisers to create audiences that exclude certain protected classes.
%
%In the U.S., doing so for ads for housing~\cite{FairHousingActDiscrimination}, credit~\cite{EqualCreditOpportunityAct}, or employment~\cite{AgeDiscriminationinEmploymentAct} may constitute illegal discrimination.
%
In March 2018, the NFHA and others sued Facebook for allowing landlords and real estate brokers to exclude members of protected groups from receiving housing ads~\cite{FacebookCivilLawsuit}.
The lawsuit was settled in March 2019, and Facebook agreed to make a number of changes to its ad targeting tools. 
%
% Relevant here, Facebook agreed to change how Lookalike Audiences (LAL) works when used with housing, credit, and employment (HEC) ads such that age, gender, relationship status, religious views, school, political views, interests, or zip code would not be considered in the search of ``similar'' users.
% \if \icwsm~\cite{FacebookNFHASettlement}:\fi 
% %
% \begin{displayquote}
% 5. Lookalike Audience (``LAL''): In the HEC Flow, LAL tool and marketing will be modified as follows:
% \smallskip
% %~\\
% (a) LAL tool may consider the following user profile fields: country, region, profession and field of study. LAL tool will not consider the following user profile fields: age, gender, relationship status, religious views, school, political views, interested in, or zip code.
% \end{displayquote}
%
Facebook now refers to this modified Lookalike Audiences tool as {\em Special Ad Audiences}.\if \icwsm~\cite{FacebookSpecialAdAudience}. \fi 
%
% Facebook says that Special Ad Audiences ``will create an audience based on similarities in online behavior and activity but that does not use certain categories, including age, gender, ZIP code or other similar categories.''
%\if \icwsm~\cite{FacebookSpecialAdCategory}.\fi  
%\begin{displayquote}
%
%[Special Ad Audiences] will create an audience based on similarities in online behavior and activity but %that does not use certain categories, including age, gender, ZIP code or other similar categories.
%\end{displayquote}
%

From an advertiser's perspective, Special Ad Audiences are created in the same manner as Lookalike Audiences (i.e., based on a source Custom audience).
The minimum size for both types of these algorithmically generated audiences is 1\% of the population of the target location, regardless of the size of the source audience.
In case of the US that means that the algorithm outputs audiences of 2.3 million users.
%
% A screenshot of Facebook's interface for creating Lookalike Audiences is shown in Figure~\ref{fig:creation} (left), and for Special Ad Audiences in Figure~\ref{fig:creation} (right).
%

% \textcolor{red}{ Repeated from earlier: 
% The co-existence of Lookalike Audiences and Special Ad Audiences offers a unique opportunity: it allows us to compare how complex machine learning systems perform on large-scale, real-world data when one does and does not include protected attributes.
% %
% Put simply, do Special Ad Audiences actually produce less discriminatory audiences than Lookalike Audiences?}

\subsection{Related work}

% We briefly overview related work on studying and mitigating algorithmic bias, as well as Facebook's advertising platform.

% \para{Algorithmic bias}
% \para{Procedural fairness}
% Concerns over bias in algorithms have galvanized a growing research community.
% %
% This community has developed a number of approaches to {\em algorithmic auditing}~\cite{sandvig-2014-auditing}, a process of seeking to understand an algorithm's inputs, outputs, and potential for discriminatory effects.
% %
% Researchers have successfully studied a variety of widely deployed algorithmic systems including face-recognition systems~\cite{buolamwini-2018-gender}, e-commerce sites~\cite{hannak-2014-ecommerce}, search engines~\cite{hannak-2013-filterbubbles,kulshrestha-2017-quantifying,diakopoulos-2018-vote,robertson-2018-auditing,kay-2015-unequal}, job seeking sites~\cite{hannak-2017-bias,chen-2018-chi}, online translation services~\cite{bolukbasi-2016-man}, housing sites~\cite{asplund-2020-auditing}, image-tagging APIs~\cite{kyriakou-2019-fairness}, or health-management~\cite{obermeyer-2019-dissecting}.
% %
% A number of proposals have been put forward to mitigate the potential algorithmic biases; we refer the reader to the survey by Mehrabi~\etal~\cite{mehrabi-2019-survey} for a more in-depth treatment.
% %
% We highlight a few works most closely related to our topic of measurement.
%
Greenberg distinguishes two kinds of fairness concerns, {\em distributive} and {\em procedural}~\cite{greenberg-1987-taxonomy}.
The former aims to assure balanced outcomes, whereas the latter focuses on the process itself.
Elimination of sensitive features, for example sex or race, from an algorithm's input (as with Special Ad Audiences) falls into the {\em procedural} category.
Such approach in the legal context is also referred to as \textit{anti-classification} and it is encoded in the current standards~\cite{davies-2018-measure}. 
However, scholars and researchers have for decades critiqued this so-called ``colorblind'' approach to addressing historical inequality and discrimination~\cite{bonilla2006racism}. 
Legal scholar Destiny Perry argues that ``(1) colorblindness is, under most circumstances, undesirable given its recently discovered negative outcomes, particularly for the very groups or individuals it is meant to protect; (2) true colorblindness is unrealistic given the psychological salience of race; and (3) race consciousness in the law is necessary to ensure equal treatment of racial groups in regulated domains such as housing, education, and employment~\cite{peery2011colorblind}.''
In the context of sentencing and mass incarceration Traci Schlesinger concludes that ``in the post-civil rights era, racial disparities are primarily produced and maintained by colorblind policies and practices~\cite{schlesinger2011failure}.''
Similar arguments have been made in the context of housing discrimination and a range of other domains~\cite{anderson2004colorblind}.

Previous work in statistics and machine learning indicated that, in general, removing sensitive features does not reliably achieve fairness for a number of reasons. 
First, certain features might serve as close proxies for the sensitive information.
For example, due to housing segregation a person's zip-code can be predictive of their race.
%
% In fact, as part of the settlement Facebook, the zip-code is also removed for creation of Special Ad Audiences, specifically because its correlation with race.
%
Second, the removed information might be {\em redundantly encoded} by non-sensitive features or their combinations.
It will then be reconstructed by the model if it is pertinent to the prediction task~\cite{dwork-2012-fairness,yeom-2018-hunting,CfpbProxy}.
%or even if it just appears pertinent only because the data used to train the model encodes historical discriminatory treatment~\cite{zafar-2015-fairness}. 
One such example is the fiasco of Amazon's hiring algorithm~\cite{goodman-2018-amazon}. 
%
% Though unaware of the applicants' gender, the model learned to replicate historical bias against candidates from all-women's colleges and in favor of candidates that whose letters included words like ``executed'' and ``captured'', predominantly used by men.
%
Third, there are cases in which only certain intersections of values of otherwise non-sensitive features are to be protected~\cite{pedreschi-2008-discrimination}.
Finally, even if none of the features or their combinations are unfair, their predictive performance might differ across sub-populations.
In an effort to minimize the total error, the classifier will fit the majority group better than the minority~\cite{sapiezynski-2017-imbalance,chen-2018-my}.
Taken together, these prior works paint a clear picture of process fairness, or fairness through unawareness, as insufficient to ensure fair outcomes.
Unfortunately, despite this consensus among scholars and a few high-profile failures in practice, the 2019 settlement is still based on fairness through unawareness. 
In this article we investigate whether this particular implementation is closer to achieving the goal of fairness. 

% %
% Grgi\'c-Hla\v{c}a~\etal~\cite{grgic-2018-beyond} propose a framework which relies on human moral judgments to determine which features are fair to use.
% %
% They point out that while people can accurately judge relevance and privacy aspects of a feature in decision making, they tend to fail at predicting the impact that feature might have on the decision outcomes.
% %
% Specifically, certain features might appear fair to human judges even though they are correlated with sensitive features.
%

Regardless of the particular approach to ML fairness, focusing on particular algorithms can be too narrow of a problem definition. 
Real-world algorithmic systems are often composed of multiple subsystems and can be discriminatory as a whole, even if built from a series of fair algorithms~\cite{dwork-2018-fairness}.
They need to be modeled along with the other components of the {\em socio-technical} systems they are embedded in~\cite{selbst-2019-fairness}.
The burden of these investigations lies on independent researchers and auditors since the companies who operate these algorithms might not be incentivized to measure and address the externalities they cause~\cite{overdorf-2018-questioning}.

% \para{Facebook's advertising platform}
% Facebook runs one of the world's most powerful advertising platforms, and has been the object of study for a number of research projects.
% %
% Prior work has demonstrated that Facebook was using PII provided for security features (e.g., two-factor authentication) was used to allow advertisers to target users with ads~\cite{venkatadri-2019-pii}, that Custom Audiences can be used to leak user's PII~\cite{venkatadri-2018-targeting}, that Facebook's ad targeting options offer a variety of mechanisms to create discriminatory audiences~\cite{speicher-2018-targeted}, that political advertisers on Facebook with higher budgets target people using more privacy sensitive features~\cite{ghosh-2019-facebook} and that Facebook's ad delivery system {\em itself} may introduce unwanted biases when deciding which users should be presented with life opportunity~\cite{lambrecht-2018-algorithmic,ali-2019-discrimination} and political ads~\cite{ali-2019-arbiters}.

%
%In this paper, we do not aim to study Facebook's implementation of Special Ad Audiences itself, but rather, use its existence as an opportunity to study the ability of algorithms to use protected features when they are not provided them directly.

%!TEX root = main.tex

\section{Methodology}\label{sec:methodology}
In this work we attempt to measure the audience skews in terms of gender, age, race, and political views. Facebook Ad Manager reports the gender and age distribution of the audiences that received each ad, but it does not report the information about the race or political views of these audiences. We therefore apply two different approaches to creating the audiences and measuring the effects.

\subsection{Timing}
The 2019 settlement~\cite{FacebookNFHASettlement} stipulated that the updated ad creation flow for special categories be implemented by September 30, 2019. 
All of our ads were created and run between October 20, 2019 and December 15, 2019, leaving Facebook ample time after the implementation deadline.

\subsection{Measuring skews by gender and age}
To measure the makeup of a target audience by gender and age, we create and run actual ads and then we use the Facebook Ad Manager API to record how they are delivered.
For these experiments, we need to provide an {\em ad creative} (consisting of the ad text, headline, image, and destination URL).
Since the ad content influences the delivery~\cite{ali-2019-discrimination}, we chose to use the same creative for all ads, unless otherwise noted: a generic ad for Google Web Search, which has basic text (``Search the web for information'') and a link to Google.
We found that Facebook does not verify that an ad that is self-reported by an advertiser as a housing, credit, or employment ad is, in fact, such an ad.
On the other hand, Facebook does automatically classify housing, credit, or employment ads as such even if the advertisers chooses not to disclose that information.
Thus, the only way for us to run the same ad creative using both Lookalike and Special Ad audiences was to run a neutral ad that would not trigger the automatic classification.

\para{Creating audiences}
Recall that our goal is to measure whether Special Ad Audiences produce significantly less biased audiences than Lookalike Audiences.
We therefore need to generate source audiences with controlled and known bias, from which we can create a Lookalike and a Special Ad audience.
We replicate the approach from prior work~\cite{ali-2019-discrimination}, relying on publicly available voter records from New York and North Carolina.
These records include registered voters' gender, age, location (address), and (only in North Carolina) race.

Thus, for each demographic feature we wish to study, we first create a Custom audience based on the voter records (which we treat as ground truth).
For example, when studying gender, we select a subset of the voters who are listed as female and use that list to create a Custom audience.
We use each biased Custom audience to create both a Lookalike audience and a Special Ad audience, selecting users in the U.S. and choosing the smallest size option	 (1\% of the population).

%If Special Ad Audiences were working as intended, we would hope that the Special Ad Audience we created from a Custom Audience of (say) all women would be {\em significantly less} biased than the Lookalike Audience created from the same source (as the Special Ad Audience was created without having access to Facebook users' gender). 
%
%However, if the two ads both show significant bias, and that significant bias is similar, it suggests that the ML system that creates Special Ad Audiences is able to infer protected attributes (gender) despite not being given them explicitly.

\para{Data collection}
Once the ads are running we use Facebook's Ad Manager tool to collect information about demographics of the audiences that Facebook shows our ads to, broken down by age group, gender, and the intersections of these two characteristics.

\para{Calculating and comparing gender skew}
The Ad Manager tool reports gender of each user as either female, male, or unknown. 
The unknown gender might refer to users who choose to self-report their gender as falling outside of the binary, or those who did not provide their gender.
We note that in all experiments there is no more than 1\% of such users, and report the observed gender bias as the fraction of men $\hat{p}$ in the reached audience.
We also calculate the upper and lower 99\% confidence intervals ($U.L$ and $L.L$, respectively) around this fraction $\hat{p}$ using the method presented by Agresti and Coull~\cite{agresti-1998-approximate}:
\begin{equation}
\begin{aligned}
L.L. &= \frac{\hat{p}+\frac{z^2_{\alpha/{2}}}{2n}-z_{\alpha/2}\sqrt{\frac{\hat{p}(1-\hat{p})}{n}+\frac{{z^2_{\alpha/2}}}{4n^2}}}{1+{z^2_{\alpha/2}}/n},\\
U.L. &= \frac{\hat{p}+\frac{z^2_{\alpha/{2}}}{2n}+z_{\alpha/2}\sqrt{\frac{\hat{p}(1-\hat{p})}{n}+\frac{{z^2_{\alpha/2}}}{4n^2}}}{1+{z^2_{\alpha/2}}/n},
\end{aligned}
\label{eq:single_99}
\end{equation}
We set $z_{\alpha/2}=2.576$, corresponding to the 99\% interval.

Finally, we verify whether the difference between fractions observed for Lookalike and Special Audiences is statistically significant using the difference of proportion test:
\begin{equation}
\Delta_{p_{LS}}=(\hat{p_L}-\hat{p_S})\pm z_{\alpha/2}\sqrt{\frac{\hat{p_L}(1-\hat{p_L})}{n_L}+\frac{\hat{p_S}(1-\hat{p_S})}{n_S}},
\label{eq:frac_diff}
\end{equation}
where $\hat{p_L}$ and $\hat{p_S}$ are the fractions of men who saw the ad in the Lookalike and Special audiences, $n_L$ and $n_S$ are number of people reached in each of these audiences.
Because we are testing the significance in seven experiments (one for each input proportion), we apply the Bonferroni correction for multiple hypotheses testing.
We do so by setting $z_{\alpha/2}$ to $3.189$, corresponding to Bonferroni corrected $p_{val}=0.01/7\approx0.00143$.
If the confidence interval includes 0, we cannot reject the hypothesis that the fraction of men is the same in the two audiences and thus the result is not statistically significant.

\para{Calculating and comparing the age skew}
Age of the users who were shown each ad is reported in groups: <18, 18-24, 35-44, 45-54, 55-64, and 65+. 
We calculate the mean age and the confidence intervals around it using formulas specific to grouped data.
First, we compute the mid-point $M_i$ for each age range $i$, 
\begin{equation}
M_i=\frac{x_{min_i} + x_{max_i}}{2}
\end{equation}
Next, we find the mean age $\mu$
\begin{equation}
\mu=\frac{\sum_i{(M_i \times F_i)}}{\sum_{i}{F_i}},
\end{equation}
where $F_i$ is the number of audience members in the age group $i$.
We then compute the standard deviation around that mean
\begin{equation}
\sigma = \sqrt{\frac{\sum_i{(F_i\times M_i^2)}-(n \times \mu^2)}{n-1}}
\end{equation}
and the corresponding standard error
\begin{equation}
SE=\frac{\sigma}{\sqrt{n}}
\end{equation}
Presented upper and lower confidence intervals correspond to
\begin{equation}
\begin{aligned}
U.L. = \mu + k\times SE,\\ 
L.L. = \mu - k\times SE
\end{aligned}
\end{equation}
respectively, and $k$ is set to $2.576$.

Finally, we verify whether the difference in mean ages between the Lookalike and Special audiences is statistically significant.
To achieve that, we compute the standard error of the difference
\begin{equation}
SE_{LS} = \sqrt{\frac{\sigma_L^2}{n_L}+\frac{\sigma_S^2}{n_S}}
\end{equation} 
and the 99\% confidence interval around the difference between mean ages:
\begin{equation}
\Delta_{\mu_{LS}}=\mu_L-\mu_s\pm z_{\alpha/2} \times \sqrt{\frac{\sigma_L^2}{n_L}+\frac{\sigma_S^2}{n_S}}
\end{equation}
We apply the Bonferroni correction for six tests and use the $z_{\alpha/2}$ set to $3.143$. 
If the confidence interval includes 0, we cannot reject the hypothesis that the mean age is the same in the two audiences and thus the difference is not statistically significant.

\subsection{Measuring racial skews}
When measuring racial skew in the audiences we are unable to re-use the same methodology for age and gender, which relied on Facebook's {\em ad delivery} statistics.
Instead, we develop an alternative methodology that relies on {\em estimated daily results}\if \icwsm~\cite{FacebookReachDiff}\fi -- Facebook's estimate of the number of users matching the advertiser's targeting criteria that can be reached daily within the specified budget. 
We set the daily budget to the maximum allowed value (\$1M) to best approximate the total number of users that match the targeting criteria.
Facebook returns these values as a range (e.g., ``12,100 -- 20,400 users''); throughout this procedure, we always use the lower value.\footnote{We used the midpoint and the upper value and found similar results.}
The procedure has only two steps: audience creation and targeting. 
It does not involve running any ads and observing the skew in delivery, and it is entirely based on the estimates on audience sizes provided by Facebook at the ad targeting step.

We note that ours is not the first use of these estimates to infer the number of users that match different criteria. For example, Garcia \etal used them to estimate the gender inequality across the globe~\cite{garcia-2018-analyzing}, while Fatehkia \etal found they are highly predictive of a range of other social indicators~\cite{fatehkia2020mapping}. 
%
% First, we upload source audiences that consist of users of one race per audience and then create Lookalike and Special Ad audiences from them.
% %
% We also upload larger {\em reference} audiences, also one for each race, which we do not use to generate Lookalike or Special Ad audiences.
% %
% These reference audiences do not contain any users included in the source audiences for Lookalike and Special ad audience creation.
%
% As the second step, we leverage the estimated daily results to approximate the overlap between algorithmically generated audiences (with unknown race) and our ground-truth reference audiences (with known race).
%

%

\para{Audience Creation}
We start with the publicly available voter records from North Carolina, in which the voters self-report their race and ethnicity.
We focus on two groups: Non-Hispanic Black and Non-Hispanic white.
For each group, we create two independent Custom audiences: one list of 10,000 randomly selected users with that race, and one list of 900,000 randomly selected users with that race.
The latter audience does not contain any individuals already selected for the first list, and will be refered to as the \textit{reference} audience.

We refer to these as \texttt{w\_10k} and  \texttt{w\_900k} (white audiences) and  \texttt{b\_10k} and  \texttt{b\_900k} (Black audiences).
We then have Facebook algorithmically generate Lookalike and Special Ad audiences using the smaller Custom audiences as input.
We refer to the resulting audiences as $L_\texttt{w\_10k}$ (for the Lookalike audience based on \texttt{w\_10k}), $S_\texttt{w\_10k}$ (for the Special Ad audience), $L_\texttt{b\_10k}$, and $S_\texttt{b\_10k}$.

\para{Targeting}
The goal of this step is to find the overlaps between the audiences with unknown race generated by the algorithms and the reference Custom audiences that we provided (with known race).
Then we can say there is a race bias in the white Lookalike audience $L_\texttt{w\_10k}$ if the overlap between it and a white reference audience \texttt{w\_900k} is higher than the overlap between it and a Black reference audience \texttt{b\_900k} (and vice versa for an audience generated from a Black source audience).
We also perform these overlap comparisons for Special Ad audiences to measure whether this effect persists despite removing sensitive features from the algorithm.
%
% From previous work we know that race or correlated features are used to generate Lookalike audiences~\cite{speicher-2018-targeted}. 
%
% This leads us to expect that the overlap between the large white audience \texttt{w\_900k} and the audience generated from a white sample $L_\texttt{w\_10k}$ would be higher than between \texttt{w\_900k} and $L_\texttt{b\_10k}$ (and vice versa, the overlap between \texttt{b\_900k} and $L_\texttt{b\_10k}$ would be higher than \texttt{b\_900k} and $L_\texttt{w\_10k}$).
%

% We now use the ad targeting interface to obtain such estimates.
% %
% To do so, we begin the ad creation process and set our budget is the maximal value allowed (\$1M/day).
% %
% We also specify that we only target users in North Carolina.

Our method relies on the fact that Facebook allows advertisers not only to specify which audiences to {\em include} in the targeting, but also which to {\em exclude}.
Suppose we wish to obtain an estimate of the fraction of white users in $L_\texttt{w\_10k}$.
To do so, we first target the reference white audience \texttt{w\_900k} audience and record the potential daily reach (e.g., 81,000).
We then target $L_\texttt{w\_10k}$ and record the potential daily reach (e.g., 397,000).
Finally, we target $L_\texttt{w\_10k}$ and {\em exclude} the \texttt{w\_900k} audience, and record the potential daily reach (e.g., 360,000).
Now, we can observe that excluding \texttt{w\_900k} from $L_\texttt{w\_10k}$ caused the potential daily reach to drop by 37,000, indicating that approximately 46\% (37,000/81,000) of \texttt{w\_900k} were present in $L_\texttt{w\_10k}$.
We can then repeat the process with excluding \texttt{b\_900k}, and measure the fraction of the reference Black audience that is present in $L_\texttt{w\_10k}$.
By comparing the fraction of \texttt{w\_900k} and \texttt{b\_900k} that are present in $L_\texttt{w\_10k}$, we obtain an estimate of the racial bias of $L_\texttt{w\_10k}$.

% \paragraph{Robustness}
% We further confirm that the presented results are robust to the random selection of which users within each race to include in the source audiences from which Lookalike and Special Ad audiences are created.
% %
% To this end, we repeat the described process with 20 non-overlapping \texttt{b\_10k} and 20 non-overlapping \texttt{w\_10k} audiences, again created using the publicly available voter records from North Carolina.
% %
% We create a Lookalike and a Special Ad audience for each of those and then compute the overlap with large \texttt{b\_700k} and \texttt{b\_700k} audiences.
% %
% Knowing the observed distributions of overlaps, we can use the two-sample Kolmogorov–Smirnov test to compare them and answer (1) whether the generated audiences are systematically skewed along racial lines and (2) whether the Special Ad audiences are less biased than Lookalike Audiences.

\para{Measuring political skews} To measure political skews we follow the exact same method as with measuring racial skews, but rather than constructing the audiences based on their reported race, we use their registered political affiliation as Democratic or Republican voters.

\para{Limitations}
Unlike in our experiments with gender and age, here we do not know the race of a vast majority of the audience.
The Lookalike and Special Ad audiences that Facebook creates consist mostly of people who appear not to be in our voter records. There are multiple reasons for why this might be the case: (1) we only looked and single race, non-Hispanic white and Black voters, excluding all Hispanic voters, as well as those of other races, and multi-racial; (2) the users in the created audiences and could be located in other states - while creating lookalike and special audiences the advertiser can only select the country where those audiences would be located.
Thus, the results we present in this section only refer to the fraction of voters with known race who are included in each Lookalike and Special Ad audience, not the racial composition of these audiences overall. %, as further emphasized in Figure~\ref{fig:race}.
Still, these estimates do give us a small window into the makeup of the Lookalike and Special Ad audiences.

%!TEX root = main.tex

\section{Results}\label{sec:results}

We now present our experiments and analyze whether Lookalike and Special Ad Audiences show similar levels of skew.  

\begin{figure}[t!]
	\includegraphics[width=1\linewidth]{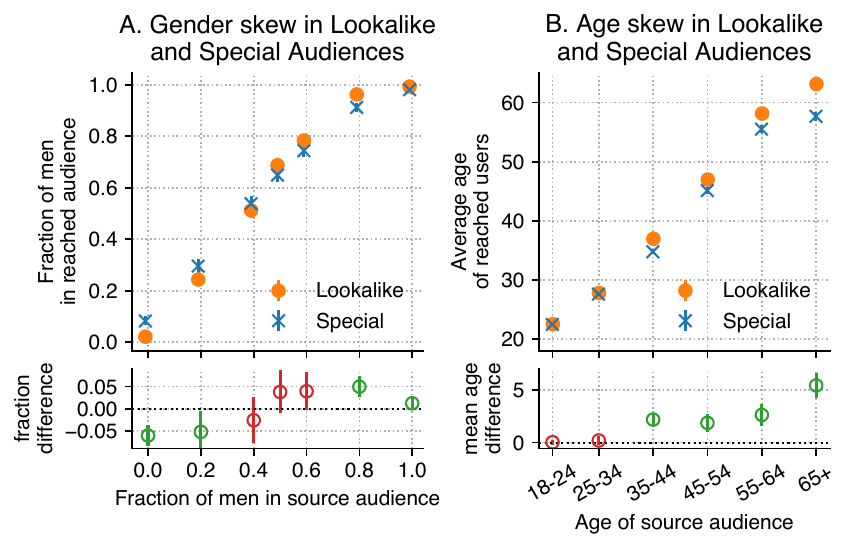}
	\caption{A. Gender breakdown of ad delivery to Lookalike and Special Ad audiences created from the source audiences with varying fraction of male users. The Special Ad audiences replicate the skew to a large extent. B. Age breakdown of ad delivery to Lookalike and Special Ad audiences created from source audiences with varying age brackets. Both Lookalike and Special Ad audiences follow the age distribution of the source audiences, but the latter shows a decrease of mean age by up to six years in the 65+ group.}

	\label{fig:gender_age}
	\vspace{-1em}
\end{figure}
%
%We first examine whether Lookalike and Special Ad Audiences can be biased along the lines of gender and age, which are straightforward to measure as Facebook provides delivery statistics in the advertiser interface.
%
%Next, we focus on race and political views, using a different methodology.
%
%Finally, we show the real-world implications of these experiments using a series of employment and credit ads.

% \newcolumntype{R}{>{\raggedleft\arraybackslash}X}

% \begin{table}[t]
% \begin{tabularx}{0.485\textwidth}{Xc|r|r}
% \small
% {\bf Source} & {\bf Type} & {\bf Male \%} & {\bf Female \%} \\
% \hline
% \multirow{2}{*}{100\% Male} & Lookalike & 95.5 & 4.5 \\
% & Special & 89.8 & 10.2 \\
% \hline
% \multirow{2}{*}{50\% Male, 50\% Female} & Lookalike & 52.5 & 47.5 \\
% & Special & 55.5 & 44.5 \\
% \hline
% \multirow{2}{*}{100\% Female} & Lookalike & 3.9 & 96.1 \\
% & Special & 8.8 & 91.2 \\
% \end{tabularx}
% \caption{Breakdown in gender of delivery audience of ads to Lookalike and Special Ad Audiences created from the same source audience, using the same ad creative.  We can observe extremely similar levels of bias, despite the lack of gender as an input to Special Ad Audiences.}\label{table:gender}
% \end{table}

\subsection{Gender and age}

We begin by focusing on gender, creating seven Custom audiences based on New York voter records.
Each audience contains 10,000 voters, with varying fractions of men: 0\%, 20\%, 40\%, 50\%, 60\%, 80\%, 100\%. 
We run ads to the resulting Lookalike and Special Ad audiences, and compare the results in ad delivery as reported by Facebook's advertiser interface.

Figure~\ref{fig:gender_age}A presents a summary of the results of this experiment, and we make a number of observations.
{\em First}, we can see that each Lookalike audience clearly mirrors its source audience along gender lines: the Lookalike audience derived from a male-only source audience delivers to over 99\% men, and the the Lookalike audience derived from a female-only source audience delivers to over 97\% women.
{\em Second}, we observe a slight male bias in our delivery, relative to the source audience:  for example, the Lookalike audience derived from a source audience of 50\% men actually delivered to approximately 70\% men.
This male bias has been observed by prior work~\cite{lambrecht-2018-algorithmic,ali-2019-discrimination} and may be due to market effects or ad delivery effects (which affect both Lookalike and Special Ad audiences equally).
{\em Third}, and most importantly, when we compare the delivery of each Special Ad audience to its corresponding Lookalike audience, we observe that a similar level of skew (that in some cases is statistically indistinguishable).
For example, the Special Ad audience derived from a male-only source audiences delivers to over 95\% men, despite being created without having access to users' genders.
As emphasized the lower panel of Figure~\ref{fig:gender_age}, the Special Ad audiences do show a bit less skew when compared to the Lookalike audiences for some of the input audiences, while still carrying over most of the skew from the source audience.

We follow an analogous procedure to create six Custom Audiences, each consisting of individuals only in a specified age range.
We then create Custom and Special Ad audiences and measure whether the age skews are reproduced and present the results in Figure~\ref{fig:gender_age}B.

\begin{figure}[t!]
	\includegraphics[width=1\linewidth]{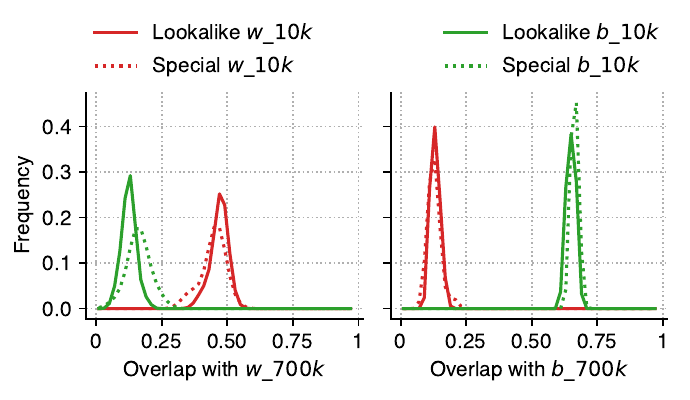}
	\caption{Both Lookalike and Special Ad audiences created from source audiences of white users containing a higher fraction of white users than Black users. Conversely, audiences created from source audiences of Black users contain a higher fraction of Black users than white users.}\label{fig:overlaps}
\end{figure}

\subsection{Race}
Next, we turn to examine the extent to which Special Ad Audiences can be biased along racial lines, in the same manner Lookalike Audiences were observed to be in past work~\cite{speicher-2018-targeted}.
% We begin by presenting Venn diagrams in Figure~\ref{fig:race} that capture the overlap between all of the audiences.
%
We summarize the overlap between the Lookalike and Special Ad audiences and the large white and Black audiences in Table~\ref{table:race}.
Focusing on the table, we can immediately observe that both the Lookalike audiences show significantly more overlap with the race of the source audience, suggesting that the makeup of the Lookalike audiences are racially biased.
For example, the Lookalike audience created from \texttt{b\_10k} contains 61\% of the active users from \texttt{b\_900k} but only 16\% of the active users from \texttt{w\_900k} (see Methods for the explanation of the audience names).
More importantly, the Special Ad audiences show a similar behavior (though as before, perhaps with slightly less of a bias).
Again, it is important to keep in mind that we can only make estimates of the fraction of  \texttt{w\_900k} and  \texttt{b\_900k} that overlap with the Lookalike and Special Ad audiences, and cannot comment on the majority of these audiences (as they likely fall outside of North Carolina).
Thus, our results are not conclusive---but only suggestive---that the overall audiences are similarly biased.
Below, we provide further robustness analysis of these results.

\subsection{Robustness}
Here, we verify that the presented results regarding race biases are robust to the random selection of seed from which Lookalike and Special Ad audiences are created.
Following the method described in Methodology, we use the two sample Kolmogorov–Smirnov test to compare the distributions of overlaps presented in Figure~\ref{fig:overlaps}.
The findings are confirmed to be robust to the particular source audience choice.
{\em First}, the racial skew observable in Lookalike audiences persists in Special Ad audiences and is statistically significant at $p_{val}=0.01$ even with the Bonferroni correction for multiple hypotheses testing,
{\em Second}, the differences between overlaps produced by Special Ad audiences and Lookalike audiences generated from the \texttt{w\_10k} custom audience \textit{are not statistically significant} -- Special Ad audiences generated from the \texttt{w\_10k} are just as biased as the corresponding Lookalike audiences.
{\em Third} the differences between overlaps produced by Special Ad audiences and Lookalike audiences generated from the \texttt{b\_10k} custom audience \textit{are small statistically significant} and this difference comes from Special Ad audiences being even more biased than Lookalike audiences.

\begin{table}[t!]
\begin{tabularx}{0.485\textwidth}{Xc|r|r}
\small
& & \multicolumn{2}{c}{\bf Percent overlap} \\
 & & {\bf Black~~~\null } & {\bf White~~~\null } \\
{\bf Source} & {\bf Type} & (\texttt{b\_900k}) & (\texttt{w\_900k}) \\
\hline
\multirow{2}{*}{100\% Black} & Lookalike ($L_\texttt{b\_10k}$) 	& 61.0 & 16.0 \\
							 & Special ($S_\texttt{b\_10k}$) & 62.3 & 12.3 \\
\hline
\multirow{2}{*}{100\% white} & Lookalike ($L_\texttt{w\_10k}$) & 16.9 & 42.0 \\
							 & Special  ($S_\texttt{w\_10k}$) & 10.4 & 35.8 \\
\end{tabularx}
\caption{Breakdown of overlap between audiences with known racial makeup and Lookalike and Special Ad audiences. While we do not know the race of the vast majority of the created audiences, we see large discrepancies in the race distribution among the known users.}\label{table:race}

\vspace{-3ex}
\end{table}

% Special:
% black rest 77 
% rep rest 81
% black 10 232 k 
% white 10 313 k 

% (29, 176, 48, 71, 0,10,0) special_black
% (69, 276, 8, 52, 0,29,0) special_white

% Lookalike:
% black 10 298 
% white 10 397 -
% (30, 239, 47, 68, 0,13,0) lookalike_black
% (64, 348, 13, 47, 0,34,0) lookalike_white

% audience	diameter	area
% 397	1.3	1.327322896
% 298	1.126305568	0.9963280178
% 313	1.154304178	1.046478757
% 232	0.9937842084	0.7756647655
% 83	0.5944114207	0.2775007566
% 77	0.5725236897	0.257440461

% \begin{figure*}
% 	\includegraphics[width=1\textwidth]{figures/aud_ads.png}
% 	\caption{Ad creatives used throughout the paper.  All of our ads linked directly to the domains shown in the ad.}\label{fig:ads}
% \end{figure*}

%\if 0
\subsection{Political views}
We next turn to measure the extent to which Lookalike and Special Ad Audiences can be biased along the lines of political views.
As with race, Facebook does not provide a breakdown of ad delivery by users' political views.
Thus, we repeat the methodology we used for race, using voter records from North Carolina and focusing on the differences in delivery to users registered as Republicans and Democrats.

% Specifically, we create source audiences of Republicans and Democrats (\texttt{r\_10k} and \texttt{d\_10k}), as well as large Republican and Democrat audiences (\texttt{r\_900k} and \texttt{d\_900k}).
% %
% We then use the source audiences to create both Lookalike audiences ($L_\texttt{r\_10k}$ and $L_\texttt{d\_10k}$) and Special Ad audiences ($S_\texttt{r\_10k}$ and $S_\texttt{d\_10k}$).
% %
% As with race, we run the same generic ad to all audiences, and examine the fraction of the large audiences that are present in the Lookalike and Special Ad audiences.

We report the results in Table~\ref{table:politics}.
We can observe a skew along political views for Lookalike audiences (for example, the Lookalike audience created from users registered as Democrats contains 51\% of \texttt{d\_900k} but only 32\% of \texttt{r\_900k}).
We can also observe that the Special Ad audiences show a skew as well, though to a somewhat lesser degree than the Lookalike audiences.
As with the race experiments, we remind the reader that we can only observe the overlap between the created audiences and the large Democrat/Republican audiences; we are unable to measure the majority of the created audiences.
However, the demonstrated skew suggests that there is a bias in the overall makeup of the created audiences.
%\fi
%\fi
% dem rest 64
% rep rest 66

% dem 10 580 k
% rep 10 580 k

\begin{table}[t!]
\begin{tabularx}{0.485\textwidth}{Xc|r|r}
\small
& & \multicolumn{2}{c}{\bf Percent overlap} \\
 & & {\bf Democrat\null } & {\bf Republican\null } \\
{\bf Source} & {\bf Type} & (\texttt{d\_900k}) & (\texttt{r\_900k}) \\
\hline
\multirow{2}{*}{Democrats} & Lookalike $L_\texttt{d\_10k}$& 51.6 & 31.8 \\
							 & Special $S_\texttt{d\_10k}$& 42.2 & 25.8 \\
\hline
\multirow{2}{*}{Republicans} & Lookalike $L_\texttt{r\_10k}$	& 28.1 & 50.0 \\
							 & Special $S_\texttt{r\_10k}$ & 25.0 & 47.0 \\
\end{tabularx}
\caption{Breakdown of overlap between source audiences with known political leaning and resulting Lookalike and Special Ad audiences. While we do not know the political leaning of the vast majority of the audiences, we see discrepancies in the distribution among the known users.}\label{table:politics}
\end{table}

\subsection{Real-world use cases}\label{subsec:realads}
%So far, we have shown that the biases of ``artificially'' created source audiences carry over to Special Ad Audiences generated based on them.
%
Next, we test a ``real-world'' use case of Special Ad Audiences.
We imagine an employer wants to use Facebook to advertise open positions to people who are similar to those already working for them.
The employer might assume that since the Special Ad Audiences algorithm is not provided with protected features as inputs, it will allow them to reach users who are similar to their current employees but without gender, age, or racial biases.
The employer would therefore upload a list of their current employees to create a Custom audience, ask Facebook to create a Special Ad audience from that, and then target job ads to the resulting Special Ad audience.

% \if \icwsm
% \begin{figure*}
%         \includegraphics[width=1\textwidth]{figures/ads_abc.png}
%         \caption{Ad creatives used throughout the paper.  All of our ads linked directly to the domains shown in the ad.}\label{fig:ads}
% \end{figure*}
% \fi

We play the role of this hypothetical employer (Facebook itself in this example, which provides employees with an \texttt{@fb.com} email address).
We then run the following experiment:  We first create a baseline audience by using randomly generated U.S.~phone numbers, 11,000 of which Facebook matched to existing users. 
We then create a Custom audience consisting of 12M generated email addresses: all 2--5 letter combinations + \texttt{@fb.com}, 11,000 of which Facebook matched to existing users; this is our audience of Facebook employees.
We create Special Ad audiences based on each of these two Custom audiences.
Finally, we run two generic job ads \if \icwsm(see Figure~\ref{fig:ads}A)\fi---each to one of these Special Ad audiences, at the same time, from the same account, with the same budget---and observe how they are delivered.

\begin{figure}[t!]
	\includegraphics[width=1\linewidth]{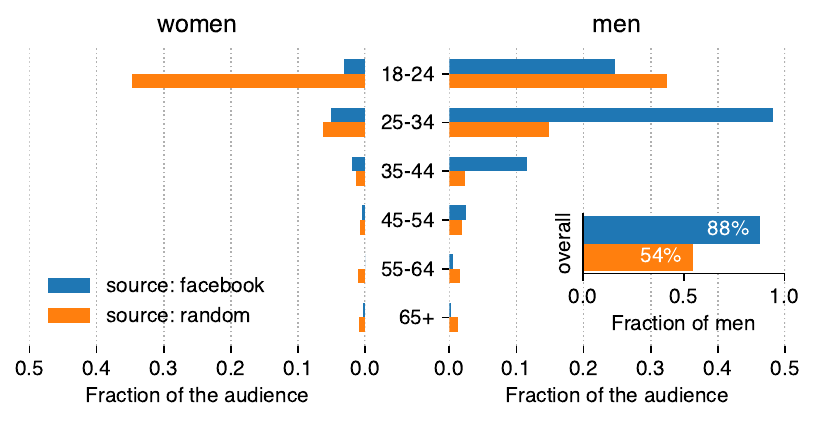}
	\caption{Gender and age breakdown of a generic job ad delivery to a Special Ad audience based on random American users (in orange) and a Special Ad audience based on Facebook employees (in blue). The audience based on Facebook employees is predominantly male and 25-34.}\label{fig:facebook_dist}
\end{figure}

Figure~\ref{fig:facebook_dist} presents the results of the experiment. 
The Special Ad audience based on Facebook employees delivers to 88\% men, compared to 54\% in the baseline case.
Further, the Special Ad audience based on Facebook employees delivers to 48\% to men aged between 25-34, compared to 15\% for the baseline audience.
Note that Facebook themselves report that the actual skew among company employees is lower, with 63\% of male employees~\cite{FacebookDiversityReport}. 
%
% Finally, 47\% of all deliveries to the Facebook Special Ad audience are to users in California, compared to 2\% in the baseline audience.\footnote{While matching based on state is not prohibited in the settlement, these numbers suggest that our method of selecting Facebook employees based on random email addresses \texttt{@fb.com} is correct.}
%
Overall, our results show that our hypothetical employer's reliance on Special Ad audiences to avoid discrimination along protected classes was misplaced:  their ad was ultimately delivered to an audience that was significantly biased along age and gender lines (and presumably reflective of Facebook's employee population).
Based on this singular experiment we cannot claim that the extent of the problem would be similar for other employers. Still, we do recommend that potential advertisers use the tool cautiously.

\subsection{Content-based skew in delivery}\label{subsec:skew}
\begin{figure}[t!]
  \includegraphics[width=1\linewidth]{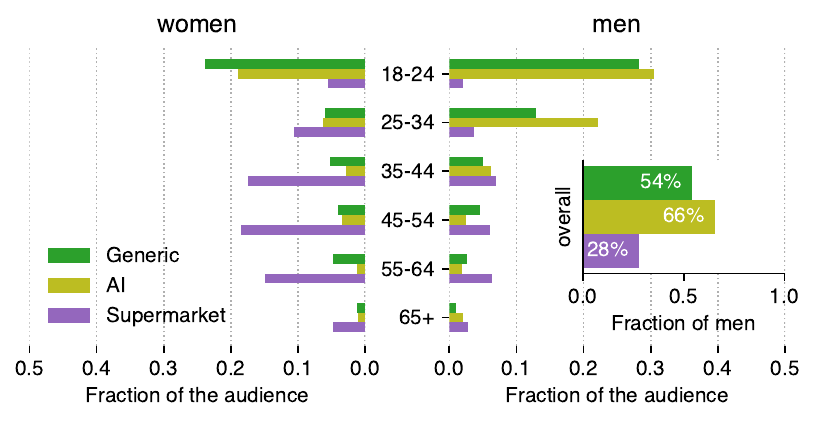}
  \caption{Gender and age breakdown of delivery of job ads to a Special Ad audience based on random American users. Facebook's delivery optimization based on the ad content can lead to large skews despite the gender and age-balanced target audience.}\label{fig:real_jobs}
\end{figure}

Previous work~\cite{ali-2019-discrimination,  imana-2021-www} demonstrated that the skew in delivery can be driven by Facebook's estimated relevance of a particular ad copy to a particular group of people.
Specifically, even when the target audience were held constant, Facebook would deliver our ads to different subpopulations: ads for supermarket jobs were shown primarily to women, while ads for jobs in lumber industry were presented mostly to men.
Here, we show that these effects persist also when using Special Ad Audiences.
%
%We re-use the Special Ad audience created from the random 11,000 users which we expect to be approximately gender- and age-balanced.
%
We run generic job ad to a Special Ad Audience created from a random set of 11,000 users along with ads for supermarket and artificial intelligence pointing to search for either keyword on \texttt{indeed.com}\if \icwsm, see Figure~\ref{fig:ads}\fi.
Figure~\ref{fig:real_jobs} shows that the different ads skew towards middle-aged women (in the case of supermarket jobs) or towards younger men (in the case of artificial intelligence jobs).

The results underline a crucial point: when designing fairness/anti-discrimination controls, one cannot just focus on one part of the {\em algorithmic} system.
Instead one must look at the whole {\em socio-technical} system, including how an algorithm is used by real people, how people adjust their behaviors in response to the algorithm, and how the algorithm adapts to people's behaviors.

\section{Legal implications}
At a high level, U.S. federal law prohibits discrimination in the marketing of housing, employment and credit opportunities. 
Our findings might have near-term legal consequences for advertisers and even Facebook itself.

A creditor, employer, or housing provider who used biased Special Ad audiences in their marketing could run afoul of the US anti-discrimination laws. 
This could be exceptionally frustrating for an advertiser who believed that Special Ad Audiences was an appropriate, legally-compliant way to target their ads.

Facebook itself could also face legal scrutiny. In the U.S., Section 230 of the Communications Act of 1934 (as amended by the Communications Decency Act, specifically 47 USC {\S} 230 {\em Protection for private blocking and screening of offensive material}) provides broad legal immunity to Internet platforms acting as publishers of third-party content. 
This immunity was a central issue in the litigation resulting in the settlement analyzed above. 
Although Facebook argued in court that advertisers are ``wholly responsible for deciding where, how, and when to publish their ads''~\cite{FacebookOnuohaMotionToDismiss}, this paper makes clear that Facebook can play a significant, opaque role by creating biased Lookalike and Special Ad audiences. 
If a court found that the operation of these tools constituted a ``material contribution'' to illegal conduct, Facebook's ad platform could lose its immunity~\cite{UpturnFacebookAmicusBrief}.

\section{Discussion}\label{sec:discussion}

We demonstrated that both Lookalike and Special Ad Audiences can create similarly biased target audiences from the same source audiences. 
We are not claiming that Facebook incorrectly implemented Special Ad Audiences, nor are we suggesting they violated the settlement.
Rather, our findings are a consequence of a complex algorithmic system at work.

Our findings have broad and narrow implications. 
Broadly, we demonstrate that simply removing demographic features from a complex algorithmic system can be insufficient to remove bias from its outputs, which is an important lesson for government and corporate policymakers. 
More specifically, we show that relative to Lookalike Audiences, Facebook's Special Ad Audiences do little to reduce demographic biases in target audiences. 
As a result, we believe Special Ad Audiences will do little to mitigate discriminatory outcomes.
% We provide a more in-depth analysis of the legal implications in the Appendix.
%

Absent any readily available algorithm-centered solutions to the presented problem, removing the Lookalike/Special Ad audience functionality as well as disabling ad delivery optimization in the sensitive contexts of housing, employment, and credit ads might be the appropriate interim approach.

\balance

\section*{Acknowledgements}
The authors thank Ava Kofman and Ariana Tobin for suggesting the experiments presented in Section~\ref{subsec:realads} as well as for going an extra mile (or two) for their ProPublica story around this work~\cite{PropublicaFacebookDiscriminates}. We also thank NaLette Brodnax for her feedback on the experimental design and Aleksandra Korolova for her comments on the manuscript. This work was funded in part by a grant from the Data Transparency Lab, NSF grants CNS-1916020 and CNS-1616234, and Mozilla Research Grant 2019H1.

\bibliographystyle{plain}
\bibliography{all}

\end{document}